**FULL TITLE:**

Addressing respiratory gating latency for accurate pulse delivery in preclinical electron FLASH irradiation on a clinical linear accelerator

**RUNNING TITLE:**

Addressing gating latency for FLASH pulse control

**AUTHOR NAMES:**

Rakesh Manjappa, PhD[1] and Jinghui Wang, PhD[1], Stavros Melemenidis, PhD[1,†], Vignesh Viswanathan PhD[1], Ramish Ashraf, PhD[1], Lawrie Skinner, PhD[1], Luis A. Soto, PhD[1], Brianna Lau, BA[1], Ryan B. Ko, BS[1,4], Rie von Eyben, MS1, Edward Graves, PhD[1], Shu-Jung Yu, PhD[1], Karl K. Bush, PhD[2], Murat Surucu, PhD[1], Erinn B. Rankin, PhD[1,3], Emil Schüler, PhD[4], Peter G. Maxim, PhD[5], Billy W. Loo Jr., MD, PhD[1]

**AFFILIATIONS:**

1. Department of Radiation Oncology, Stanford University School of Medicine, Stanford, CA 94305, USA
2. Varian Medical Systems, Palo Alto, CA 94304, USA.
3. Department of Gynecologic Oncology, Stanford University School of Medicine, Stanford, CA, 94305, USA.
4. Department of Radiation Physics, Division of Radiation Oncology, The University of Texas MD Anderson Cancer Center, Houston, TX, 77030, USA
5. Department of Radiation Oncology, University of California Irvine School of Medicine, Orange, CA 92868, USA.

**CURRENT ADDRESS:**

† Department of Radiation Oncology, Anschutz Medical Campus, University of Colorado, Aurora, CO 80045, USA

**CO-FIRST AUTHORS:**

Rakesh Manjappa and Jinghui Wang

**CO-SENIOR AUTHORS:**

Peter G. Maxim and Billy W. Loo Jr,

**CORRESPONDING SENIOR AUTHORS:**

Peter G. Maxim
Department of Radiation Oncology
University of California Irvine School of Medicine
Telephone: +1 (714) 456-8156
Email: pmaxim@hs.uci.edu

Billy W. Loo, Jr.
Department of Radiation Oncology
Stanford University School of Medicine
Telephone: +1 (650) 736-7143
Email: bwloo@stanford.edu

**AUTHOR FOR EDITORIAL CORRESPONDENCE:**

Rakesh Manjappa,
Department of Radiation Oncology
Stanford University School of Medicine
Email: rakesh9920@gmail.com

**AUTHOR CONTRIBUTIONS:**

All authors met the International Committee of Medical Journal Editors (ICMJE) criteria for authorship (https://www.icmje.org/recommendations/browse/roles-and-responsibilities/defining-the-role-of-authors-and-contributors.html).

Rakesh Manjappa and Jinghui Wang are co-first authors, each of whom is appropriately listed first when referencing this publication on their respective CVs. Peter Maxim and Billy W Loo Jr. are co-senior/co-corresponding authors.

**ACKNOWLEDGEMENTS:**

**Funding statement**

This work was supported by the Office of the Assistant Secretary of Defense for Health Affairs through the Department of Defense Ovarian Cancer Research Program under Award No. W81XWH-17-1-0042; the My Blue Dots fund; the Stanford University Department of Radiation Oncology; the Weston Havens Foundation; the Stanford University School of Medicine; the Stanford University Office of the Provost; the Wallace H. Coulter Foundation; the Cancer League; the Swedish Childhood Cancer Foundation; the Foundation BLANCEFLOR Boncompagni Ludovisi n'ee Bildt; the American Association for Cancer Research; NIH grants P01CA244091 (BWL, PGM, EEG) and R01CA266673 (ES, BWL); and philanthropic funding to the Stanford Department of Radiation Oncology.

**DISCLOSURES:**

PGM and BWL are co-founders and board members of TibaRay. BWL has received honoraria for educational lectures from Mevion. KKB is an employee of Varian Medical Systems.

**ABSTRACT**

**Background:**

Clinical linear accelerators are an accessible platform for preclinical research on the biological effects of ultra-rapid electron irradiation (FLASH). However, they are not inherently designed for the accurate pulse control required for experiments using a small number of relatively high-dose pulses, and available methods for beam control such as respiratory gating can be error-prone owing to system latency.

**Purpose:**

Here we experimentally characterize the temporal latency of the respiratory gating system for controlling beam-on and beam-off at the individual linac pulse level. We use this information to develop accurate pulse delivery methods for preclinical FLASH research.

**Methods and Materials:**

We used programmable controller boards and a relay circuit to monitor and control delivery of specific numbers of pulses through the built-in monitor chamber and respiratory gating system of a Varian Trilogy linac. We modeled system response latency as a normally distributed random variable and experimentally recorded the probability of successful pulse delivery and inhibition relative to the time of beam-on and beam-off request signals to derive the mean and standard deviation of latency times at different pulse repetition frequencies. We implemented two methods – an adaptive method using only the delivered-pulse signal, and a synchronization method additionally using the linac's internal pulse-timing signal – and characterized their performance for standard and customized pulse sequences.

**Results:**

The mean and standard deviation values of the respiratory gating latency at 60, 90 and 180 Hz pulse repetition frequency were respectively 2.0±0.8 ms, 2.1±2.8 ms, and 2.4±1.9 ms for beam-on and 1.3±0.9 ms, 1.9±2.9 ms, and 1.8±2.1 ms for beam-off. Beam-on and beam-off latencies were similar to each other, and similar across pulse repetition frequencies. Characterizing the latency parameters permitted choosing optimal timing parameters that maximized the rate of successfully delivering the desired number of pulses using both adaptive and synchronization methods, exceeding 99% at 90 Hz for both methods, and reaching 95% (adaptive) and 80% (synchronization) at 180 Hz. This also enabled successful implementation of custom pulse sequences not natively available on the system.

**Conclusions:**

We demonstrated that accounting for latency and/or using the ability to read the prior information on expected pulse timing can provide high accuracy in delivering specified numbers of pulses. This reliability is critical for accurate dose delivery in preclinical FLASH research of single fraction and especially fractionated dosing regimens. The ability to generate custom pulse sequences enables more detailed exploration of the temporal dependence of biological FLASH effects.

# INTRODUCTION

The preclinical FLASH effect (sparing of normal tissues while maintaining tumor control) associated with ultra-high dose rate (UHDR) ultra-rapid irradiation has sparked broad interest in the field because of its implications for clinical translation [1–4].

Among available radiation platforms, electron beams have been a prominent tool for FLASH research due to their practicality and compatibility with existing infrastructure [5], including dedicated research machines and conventional clinical linear accelerators (linacs) configured for preclinical electron FLASH [5–15]. A central challenge in the latter approach is beam control, as standard linac dose monitoring and pulse control are not designed for this purpose. Custom control systems are therefore required to accurately terminate the beam after a preset number of pulses is delivered, a critical need given that a single electron pulse can deliver a large proportion of the intended dose in FLASH mode (*e.g.*, >1 Gy) [8].

The accuracy of this pulse delivery is fundamentally limited by machine latency—the delay between a beam-on or beam-off command and the actual initiation or cessation of radiation [9]. For FLASH-configured clinical linacs, three beam control mechanisms are prevalent [9]:

**Respiratory Gating Switching (RGS)[10,11]:** This method repurposes the linac's existing respiratory gating interface. While highly accessible and non-invasive, it suffers from high latency (milliseconds), potentially leading to significant pulse delivery errors.

**Optocoupler Fast Circuits[12,13,16]:** These custom circuits interrupt the linac's internal trigger signals, achieving near-instantaneous beam halting with high accuracy. However, they require invasive or permanent modifications that are often restricted to decommissioned machines[16].

**Vendor-Specific Packages[14,15]:** Vendor-provided solutions offer the necessary pulse control but are limited by high cost, proprietary technology, and limited availability.

While optocoupler circuits and vendor packages offer the precision required for rigorous research, the more widely accessible RGS method is compromised by its inherent latency. To bridge this gap, we sought to optimize the RGS approach to achieve accurate UHDR dose delivery without invasive linac modification.

In this work, we first characterize the machine latency of a clinical linac using RGS and quantify its impact on pulse delivery errors. We then test practical strategies to mitigate these errors. Our aim is to provide a robust and accessible beam control solution, enabling more institutions to configure widely available clinical linacs for accurate FLASH research.

## METHODS/MATERIALS

### 2.1 FLASH linac setup

We configured a Varian Trilogy radiotherapy system (Varian Medical Systems Inc., Palo Alto, CA, USA) to perform electron FLASH irradiation, as described in recent works [10]. We used an electron beam energy of approximately **18.8 MeV** (confirmed by depth dose measurements) in the service mode for beam delivery. Linac parameters such as the radiofrequency power, pulse forming network (PFN) voltage and the electron gun grid voltage settings were configured on a dedicated electron beam control board and used for FLASH irradiations. This helped us to produce a dose per pulse ranging from **0.5 to 4 Gy** at a distance of **14.2 cm** from the scattering foil as done in a recent work [20]. Total dose was measured using Gafchromic-EBT3 film (Ashland Global Holdings Inc., Covington, KY, USA) calibrated for absolute dose, and dose per pulse was calculated from the number of pulses delivered to each film, which was counted as described below.

## 2.2 FLASH pulse control setup

We controlled the pulse delivery using a programmable controller board labelled RP1 (STEMlab 125-14, Red Pitaya, Solkan, Slovenia) and relay circuit (Bestar 500, with max. operate time of 1 ms, and max. response time of 0.5 ms), as shown in Fig. 1A. Channel 1 of the RP1 board is used to count the number of delivered pulses as detected by the internal monitor chamber ("top TP1") and impose beam hold and release through the respiratory gating system of the linac. An additional Red Pitaya board labelled RP2 (STEMlab 125-10, Red Pitaya, Solkan, Slovenia) was used as a diagnostic board for monitoring the pulse delivery. Channel 1 of this board was used to record the ion chamber signal, while the rising edge of the relay gating signal connected to Channel 2 was used as the trigger to start the acquisition. This second board was connected to a host PC through the SCPI server application and MATLAB was used for recording and visualization of these two signals as shown in Fig. 1A. A Farmer chamber (0.6 cm$^3$, PTW, Freiburg, Germany) connected to an electrometer (MAX 4000) was placed after a 10 cm solid water block downstream of the samples to be irradiated, and was used to obtain a redundant pulse count [20].

Fig. 1B shows the framing of the concept we use to understand and address the problem of the latency due to the gating. Beam-on latency is the time lag in system response to the start of the pulse delivery after the beam-on request signal (relay set high) (Fig.1B). Beam-off latency is the time lag in response to stopping of the pulse delivery after the beam-off request signal (relay set low) (Fig.1B). The number of intended pulses and pulse repetition interval determine the Intended Beam-on Duration. Offset is the time between the rising edge of the last intended pulse and the time at which the relay is set low, and can be a positive or negative value (Fig. 1C).

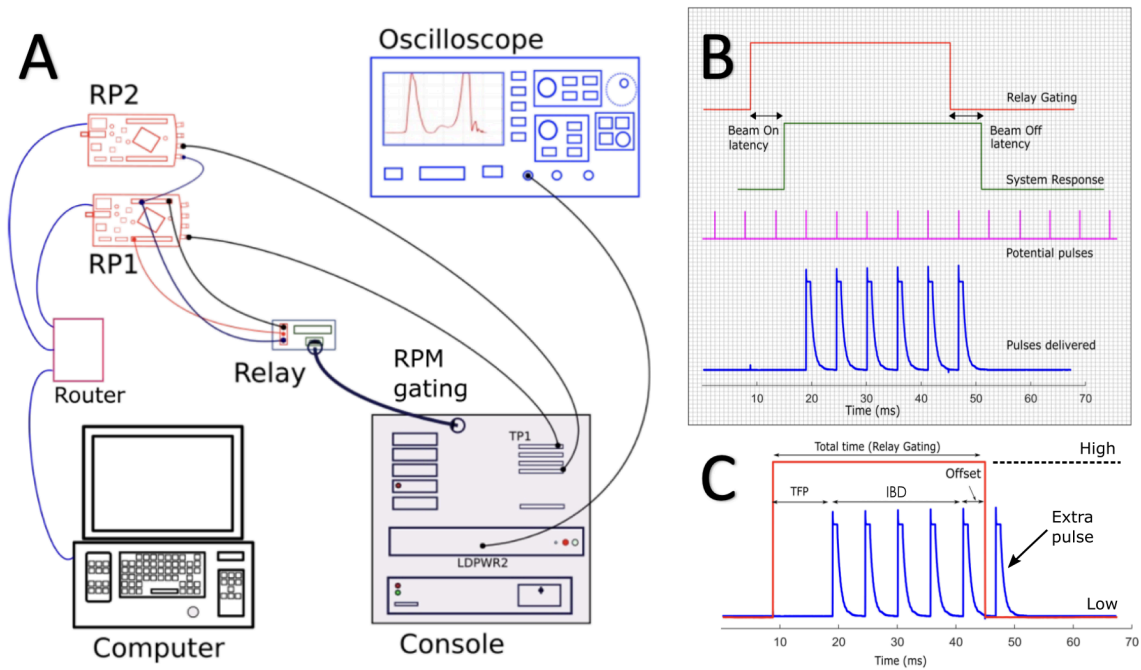

**Figure 1.**

*(A) Schematic of the electronic circuit connections used to control and monitor pulse delivery in FLASH irradiation. A microcontroller board Red Pitaya (RP1) is connected to the respiratory gating system on the linac console through a reed relay. Red Pitaya RP2 is used to measure the beam delivery with respect to relay signal. The internal ionization chamber voltage signal is read at the Top TP1 port on the linac console. The reflected signal from the console at load power R2 (LDPWR2) port is connected to Channel 1 of the oscilloscope and is used for automatic frequency control (AFC) tuning. (B) The red trace corresponds to the relay gating signal, which is the request signal we send for pulse delivery from RP1. The green trace represents the system response, which reflects the latency of the gating system with respect to request signal. The underlying potential pulses at a specific pulse repetition frequency (PRF) are shown in magenta and the blue trace is the actual pulses that get delivered and are read by both RP1 and RP2. (C) The proposed adaptive method of beam control that minimizes dose errors. The red curve is the relay signal to request beam-on and beam-off, and the blue curve is the actual delivered pulses signal. The Time to First Pulse (TFP) is measured in real time and used to decide when to set the relay low (to turn the beam off). The Intended Beam-on Duration (IBD) is determined by*

*the intended number of pulses and the pulse repetition interval, and the Offset is a positive or negative parameter used to determine the timing of the beam-off request signal (relay set low). The probability of delivering the correct number of pulses is observed to be sensitive to the value of the Offset period. This conceptual understanding helps us to characterize the latency in the system leading to errors in the number of pulses delivered.*

**2.3 Characterization of respiratory gating latency**

The shell scripting feature on the Linux operating system of the Red Pitaya was utilized to deliver multiple beams to characterize the latency in the gating system. We varied the beam-on duration by varying the Offset as a parameter. This helped us to study the dependence of accurate delivered pulse number on the Offset. For each Offset value, we recorded approximately 100 measurements (exact sample sizes are reported in §3.1). The maximum time interval that can be recorded on the Red Pitaya at the lowest time resolution, (*i.e.*, longest recording time) in one acquisition is 8.02 seconds. Hence, within each recording we delivered 25 beams with a gap of about 250 ms between each beam delivery. An example of 10 such recordings is shown in Fig. 2A.

In our initial experiments, the beam-on request signals (relay set high) were not synchronized with the underlying linac pulse timing. As such, the beam-on signals were randomly timed with respect to when potential pulses could be delivered. The timing of potential pulses (whether or not they were delivered) could be retrospectively determined from the timing of delivered pulses and knowledge of the pulse repetition interval. We determined from this whether potential pulses were missed because of beam-on latency. We recorded 1000-2000 delivery histories each at the pulse repetition frequencies of 60, 90 and 180 Hz and noted the time of the first pulse in each delivery. For each of the first three potential pulses following the beam-on signal, we assigned a score of 1 if the pulse was delivered and a score of 0 if it was not. These data were used to characterize beam-on latency. For beam-off latency characterization, we similarly scored the first three potential pulses following (prior) to the beam-off signal (relay set low), and assigned a score of 1 if the pulse was appropriately inhibited (no pulse delivered) and a score of 0 if the pulse was not inhibited (unintended pulse delivered). The scored pulses from the deliveries in Fig. 2A are shown in Fig. 2B. We sorted these pulses by time after the beam-on signal (Fig. 2C) and then binned the proportion of successful cases into time intervals (Fig. 2D). The cumulative score of a large number (thousands of such beam deliveries Fig. 2E) was used to estimate the latency parameters for beam-on and for beam-off.

For statistical analysis, we assumed the latency times for any delivery to be random with a normal distribution, and performed a probit fit with the mean and standard deviation as fitting parameters. Specifically, the probability of a pulse being delivered (or inhibited) at time t relative to the request signal was modeled as $P(t) = \Phi((t − \mu)/\sigma)$, where $\Phi$ is the standard normal cumulative distribution function, and $\mu$ and $\sigma$ are the mean and standard deviation of the latency distribution. Parameters were estimated by maximum-likelihood fitting to the binned binary outcomes, and 95% confidence intervals on $\mu$ and $\sigma$ were computed from the inverse of the Fisher information matrix.

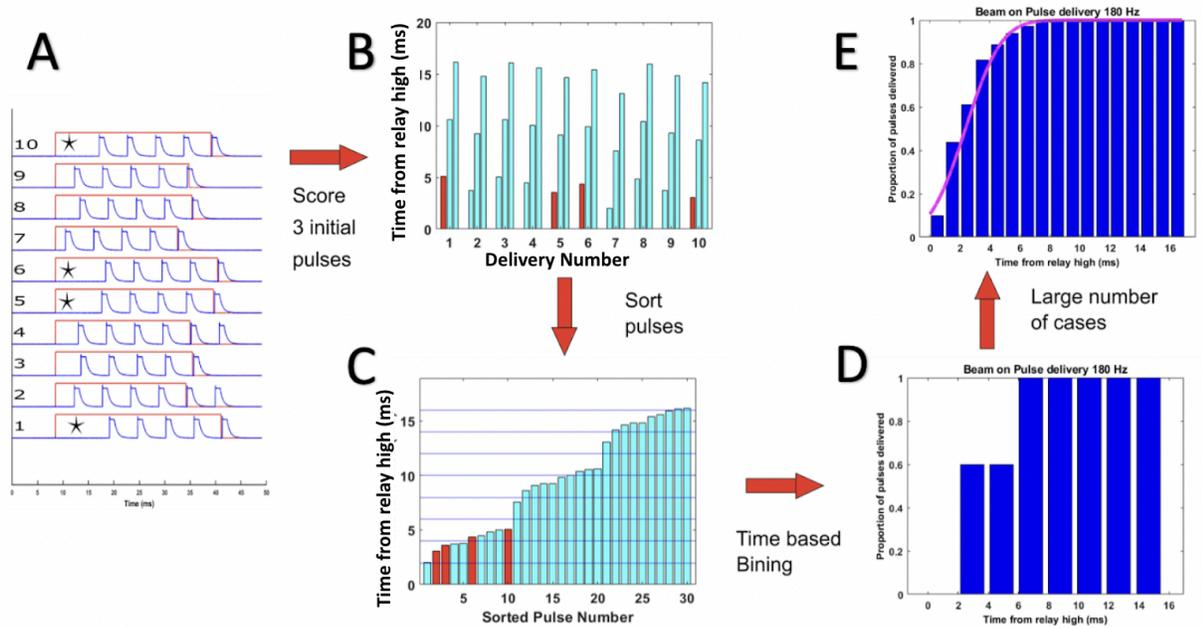

**Figure 2.** *(A) The recordings of 10 exemplary individual pulse delivery instances at 180 Hz for a delivery of 5 intended pulses. The stars indicate deliveries in which beam-on latency resulted in non-delivery of a potential pulse.*

*(B) We score the first three potential pulses for each of the deliveries to characterize the beam-on latency. If a pulse is missing at a certain expected time, we assign a score of 0 (red color) and if a pulse is delivered at an expected time then we assign a score of 1 (cyan color).*

*(C) All the pulses are sorted according to time of arrival after the beam-on request signal (relay set high) and color coded based on their scores. (D) The number of cases in each time interval are binned, and for each bin the proportion of pulses actually delivered is calculated.*

*(E) This process is repeated for a larger number of beams (~1000) to obtain a higher resolution histogram of probability of successful pulse delivery with time after the beam-on request signal (relay set high). The red curve is a probit fit to the data.*

**2.4 Monte Carlo simulation of beam delivery:**

We used the latency parameters obtained using the probit regression modeling to develop a Monte Carlo simulation of the beam delivery for comparison with experimental data. It uses an inverse distribution method for sampling. The mean and standard deviation obtained from the probit analysis are used for sampling from the normal distribution. The time from the beam-on signal (relay set high) to the beam-on system response was modeled as a random variable sampled from the normal distribution for each of the simulated beams (normrnd function in MATLAB). In the case that the value of this time was negative, we used a rejection sampling to resample from the distribution. A pulse arriving prior to the beam-on system response time would be inhibited whereas a pulse arriving after the beam-on system response time would be delivered. A step size of 1/100th the pulse period was used and the simulation recorded the total number of pulses delivered, the time of each pulse, relay high and low times. Similarly, the time from the beam-off signal (relay low) to the beam-off system response was modeled as a random variable, and pulses arriving prior to the beam-off system response time would be delivered whereas they would be inhibited after that time. For each simulated experiment, 10,000 trials were generated and the statistics compiled.

**2.5 Pulse control methods**

**Adaptive method**

Our initial experiments were performed without specifically synchronizing the beam-on request (relay set high) signal with the underlying linac pulse timing. As such, the time from the beam-on request to the first delivered pulse was variable. This Time to First Pulse (TFP) interval was measured in real time for each delivery. The total relay high time includes the TFP and the Intended Beam-on Duration (IBD), equal to the inter-pulse duration times the number of intended pulses minus one (Figure 1C). To this was added an Offset, which could be a positive or negative duration, to determine the total relay high duration (time from beam-on to beam-off request).

We varied the value of Offset to determine the optimal value to compensate for beam-off latency and ensure the highest probability of delivering the intended number of pulses. The flow chart of the algorithm used to control the beam delivery is shown in Supplementary Figure S1.

**Optimal time window method**

In some instances, detection of the time to first pulse in the adaptive method failed, in which case a default time window for the relay high signal would need to be used. Hence, we conducted optimal time window experiments. Here we varied the value of the relay high time window (from beam-on to beam-off request) to determine the maximum likelihood number of pulses delivered for each value of the time window. The time window period was varied from 40 ms to 76 ms for 90 Hz and from 10 ms to 27 ms for 180 Hz in steps of 1 ms, and 100 trials were performed at each time window. We also compared the observed results with Monte Carlo simulations. This allowed defaulting in real time to the optimal time window for the intended number of pulses in cases when the TFP failed to be detected in the adaptive method (Supplementary Figure S1).

**Synchronization method**

We hypothesized that prior knowledge of the underlying linac pulse timing would allow synchronizing the beam-on request signal to improve delivery of the intended number of pulses. We read this information from the Pulses Intended To Beam on (PLSITB) test point on the Timer board of the linac console (Fig. 1B) using channel 2 of the RP1 controller board. The time from the beam-on request signal to the next intended pulse as determined from the PLSITB signal (Time to Next intended Pulse [TNP]) was varied, and an optimal interval was identified. We refer to this technique as the synchronization method.

In these experiments, we first sought to find the optimal TNP value. In order to do this, we kept a fixed time window and vary only TNP in steps of 1 ms initially (as shown in Figure 5A), and then fine-tuned the value of TNP in steps of 0.1 ms. For each value of TNP, 100 trials were carried out to arrive at the probability of successfully obtaining the first pulse in the expected position (Figure 5B for 90 Hz and Figure 5C for 180 Hz pulse repetition frequency). As expected, the probability of success approached a maximum as TNP approaches the inter-pulse interval, 100% for 90 Hz and 86% for 180 Hz, after which it resets due to the periodic nature of pulse delivery. After obtaining the optimal TNP value, we used that in all further experiments. Then we varied the Offset to determine the optimal value to obtain the maximum rate of successful delivery of the intended number of pulses (Supplementary Table S3).

## RESULTS

### 3.1 Latency parameters

Probit analysis of the scored histories of the latency for beam-on and beam-off are plotted in Fig. 3A and 3B respectively. A sample size of 1200 measurements at 60 Hz, 2234 measurements at 90 Hz and 2048 measurements at 180 Hz were used to compute the latency parameters at each of those pulse repetition frequencies. The mean and standard deviation values of the normal distribution to fit the experimental plots at 60, 90 and 180 Hz pulse repetition frequency were respectively 2.0±0.8 ms, 2.1±2.8 ms, and 2.4±1.9 ms for beam-on latency and 1.3±0.9 ms, 1.9±2.9 ms, and 1.8±2.1 ms for beam-off latency. The beam-on and beam-off latencies were similar to each other, and similar across pulse repetition frequencies.

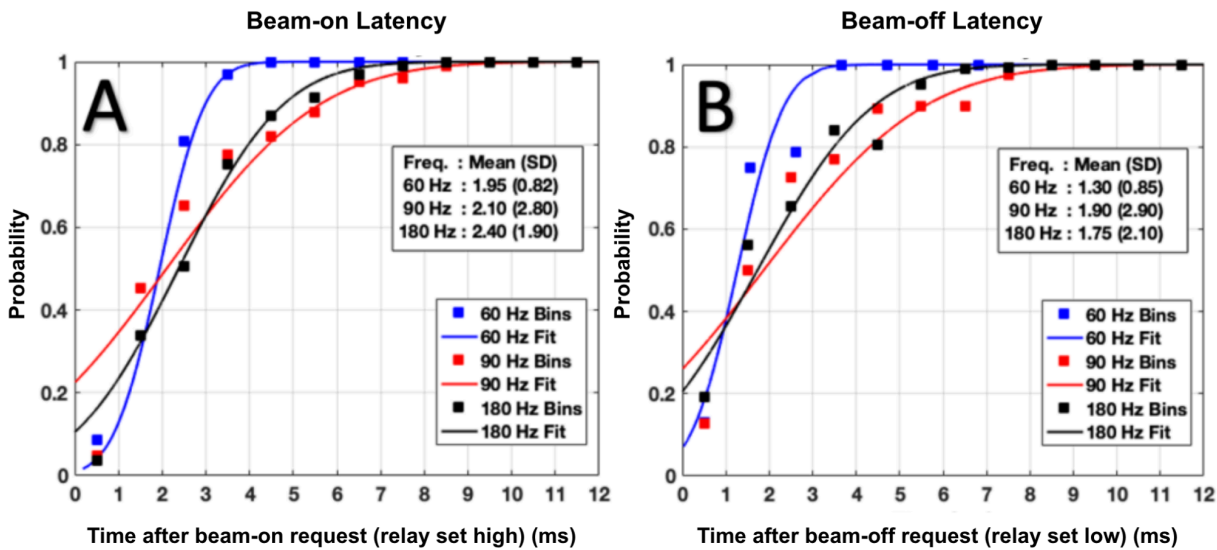

**Figure 3.** *(A) The probability of a pulse being delivered with time after the beam-on request (relay set high) signal at 60, 90 and 180 Hz pulse repetition frequency. (B) The corresponding probabilities of pulses being inhibited after the beam-off request (relay set low) signal. The experimental binned values are displayed as squares and the probit fits are plotted as solid lines. The best fit mean and standard deviation (SD) values for the latency parameters at each of the pulse repetition frequencies are indicated on the plot.*

## 3.2 Pulse control outcomes

**Adaptive method**

As can be seen in Figure 3B, there is a high probability of inhibiting a pulse if the beam-off request signal comes approximately 6 ms prior to the pulse. At a pulse repetition rate of 180 Hz, the inter-pulse interval is 5.55 ms. We found the optimal value of Offset for 180Hz to be in the range of -0.2 to -0.1 ms. Supplementary Figure S2 shows examples of deliveries using an optimal Offset of -0.16 ms, in which the number of intended pulses is delivered more reliably compared to using a non-optimal Offset of 3.83 ms. A negative Offset value means that the beam-off request signal comes slightly before the last intended pulse to provide sufficient time for inhibition of the next pulse, but close enough to the last intended pulse that the beam-off latency allows it to be delivered.

**Optimal time window method**

We simulated the number of pulses that would be delivered for fixed durations of relay high signal (from beam-on to beam-off request signals). The expected proportion of deliveries producing various numbers of pulses as a function of relay high duration at 90 Hz (3250 deliveries) and 180 Hz (1175 deliveries) pulse repetition rates are shown in Fig. 4A and 4B, respectively. We then experimentally determined the probability of successfully delivering 5 pulses at 90 Hz and 4 pulses at 180 Hz, respectively, for different time windows, compared to simulated predictions (Figure 4C and 4D). At the optimal windows in each case, the probability of delivering the intended number of pulses was approximately 80% at 90 Hz and 65% at 180 Hz. In case the Time to First Pulse could not be determined in real time in the adaptive method, defaulting to an optimal window provides the highest probability of still delivering the intended number of pulses.

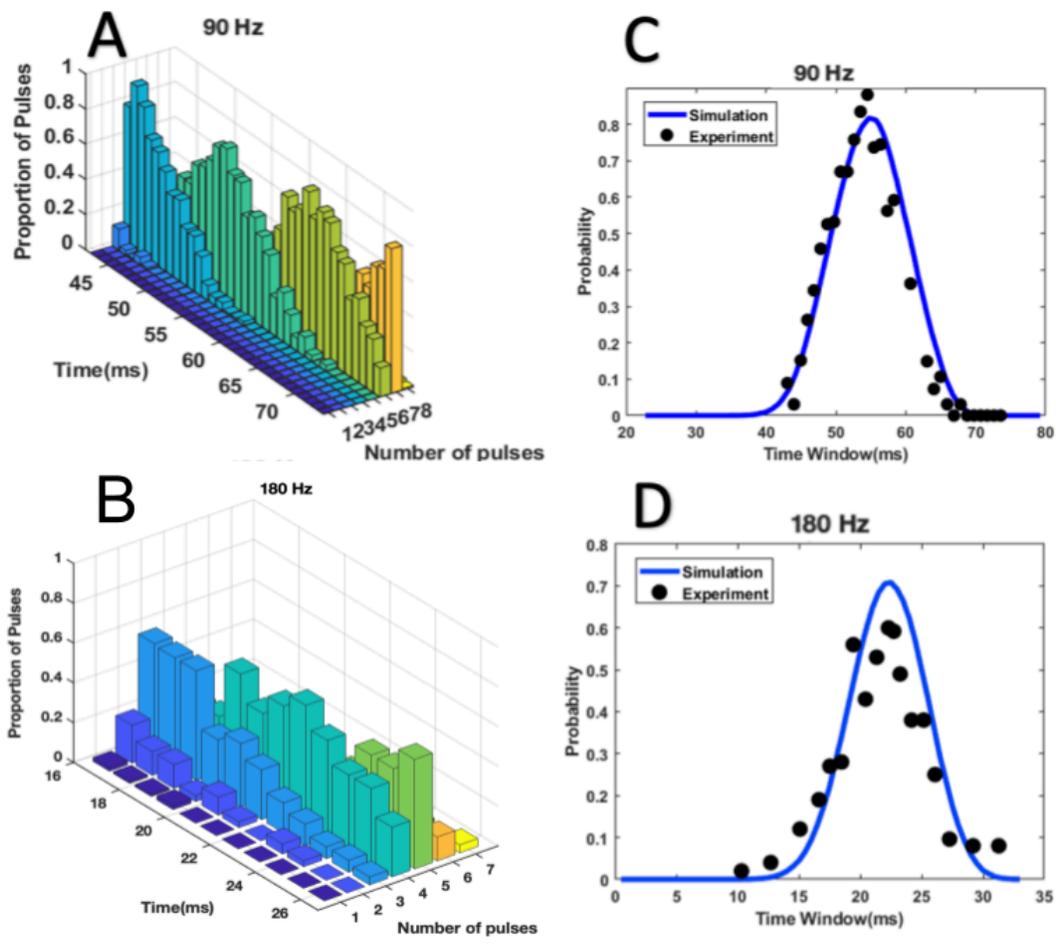

**Figure 4.** *The proportion of deliveries producing various numbers of pulses as a function of relay high duration (beam-on to beam-off request signals) at 90 Hz and 180 Hz pulse repetition rate are shown in (A) and (B), respectively. In this case, there is no synchronization of the time window with the underlying linac pulse timing or a detected first pulse delivery.*

*Experimental determination of the probability of delivering 5 intended pulses at 90 Hz pulse repetition frequency as a function of time window (relay high duration) (C) and 4 intended pulses at 180 Hz (D), respectively. The blue curves are the simulated results using the parameters obtained from the probit regression and the black circles are the experimentally measured values. At the optimal time windows, the success rate of delivering the intended number of pulses was about 80% at 90 Hz and 65% at 180 Hz. This method was used as a fallback in case the real-time determination of Time to First Pulse failed in the adaptive method. The correct number of pulses could then still be delivered most of the time even in the event of that failure mode.*

**Synchronization method**

Performance of the synchronization method is illustrated in Fig. 5A. The variation of pulses delivered with adjusting the Time to Next Intended Pulse (TNP) is studied in detail (1700 deliveries at 90 Hz and 1850 deliveries at 180 Hz). The analysis for 90 Hz and 180 Hz are shown in Fig. 5B and 5C respectively. As we approach a TNP value greater than 8 ms, combined with an optimal value of the Offset (beam-off parameter), the probability of obtaining the required number of pulses at 90 Hz approaches 1, with >**99%** success (381 out of 383 deliveries) at 90 Hz (Tables S2 & S3). For the 180 Hz case, the probability of successful delivery for a range of intended number of pulses was **81%** (851 out of 1047 deliveries) (Tables S2 & S3). Experimental beam delivery comparing the adaptive method and the synchronization method is shown in Supplementary Figure S3.

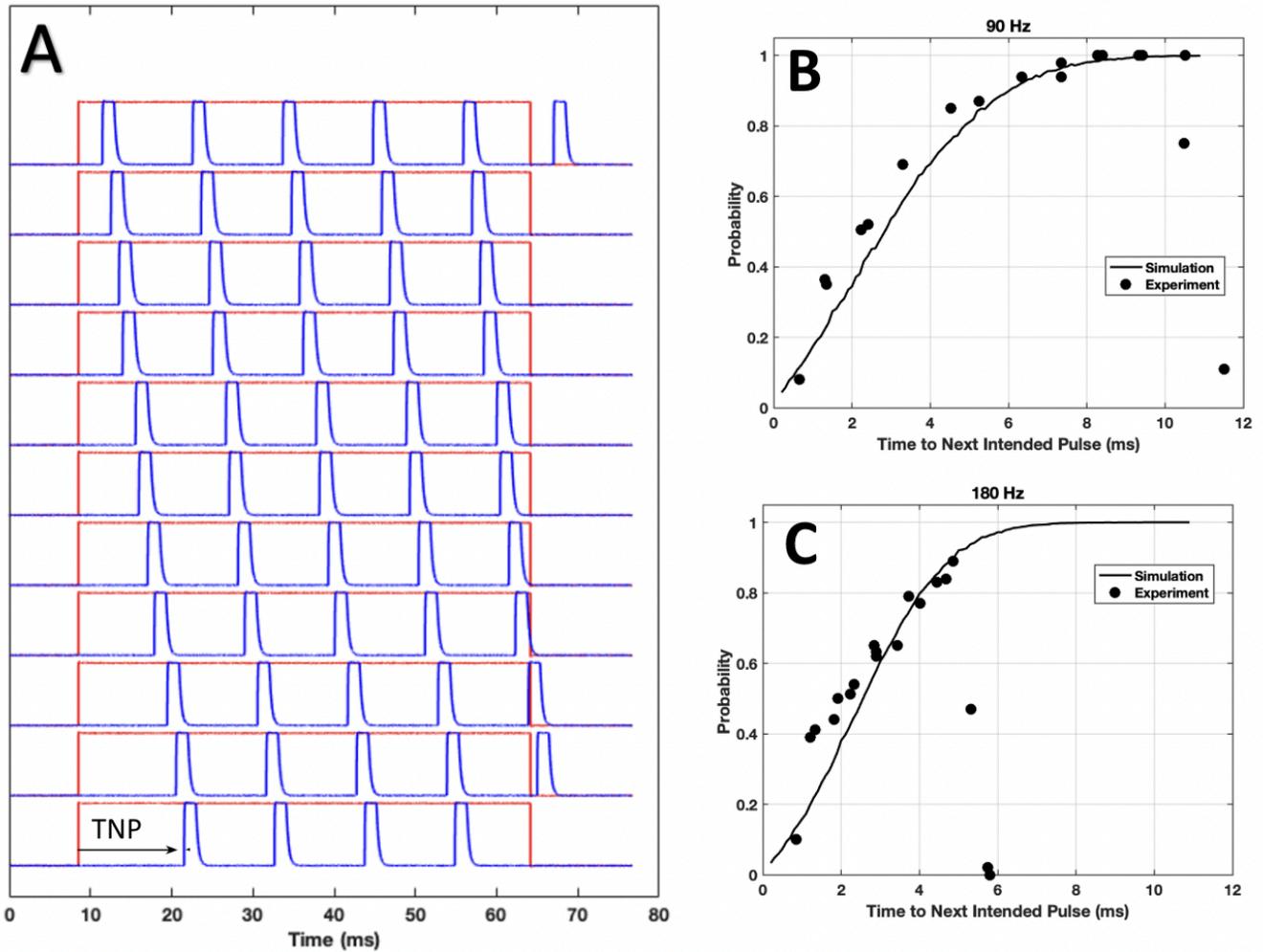

**Figure 5.** *Performance of the synchronization method of relay gating. In this method, the Pulses Intended To Beam on (PLSITB) signal was read to provide the underlying timing of linac pulses. We studied the beam delivery by varying the Time to Next intended Pulse (TNP) from the beam-on request signal (relay set high) as shown in (A) for values ranging from 0 to 11 ms in steps of 1 ms for the 90 Hz and for values ranging from 0 to 5.5 ms in steps of 0.5 ms for the 180 Hz. For each value of TNP, 100 trials were carried out to determine the probability of successfully **obtaining the first pulse in the expected position**. Plot (B) is with pulse repetition frequency 90 Hz and the plot (C) is for 180 Hz. The actual delivery (circular points) and the simulations using latency parameters obtained from the probit analysis (solid lines) are shown. The probability of success approached a maximum as TNP approaches the inter-pulse interval, 100% for 90 Hz and 86% for 180 Hz, after which it resets due to the periodic nature of pulse delivery. Results of optimizing the Offset parameter are shown in*

*Supplementary Table S3. The detailed characterization helps us understand the nature of the synchronization method and choose the optimal TNP values corresponding to the pulse repetition frequency.*

## 3.3 Customized beam delivery

We evaluated customizing the time interval between pulses (intervals not standardly available as options in the machine interface) or customizing pulse sequences such as groups of a few pulses at a time with gaps between groups, and checked the repeatability of such custom beam deliveries. Example scenarios that can be achieved using this technique are shown in Fig. 6A. The experiments that utilized these custom beams are demonstrated in Fig. 6B and 6C.

Fig. 6B highlights the case where the dose per pulse is the same (3.5 Gy per pulse) for the top and the bottom beam delivery, but the overall dose rate can be lowered to an effective 4.2 Hz which is not possible from the standard linac configuration, as the lowest is 18 Hz.

Fig. 6C demonstrates three different pulse sequences with the same overall delivery time of 277.5 ms. Among these three cases we see that the dose per pulse and intra-pulse dose rate can be varied.

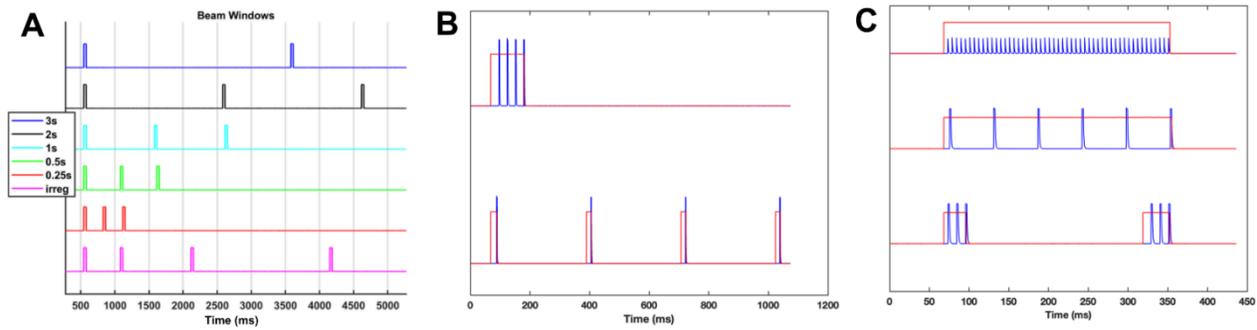

**Figure 6.** *(A) Examples showing the concept of custom beam windows with varying intervals tested for the gating signal. Using this mechanism we demonstrate intervals larger and different from the preset ones available in the standard linac configuration. (B and C). Showing the actual delivery, red is the request gating signal and blue is the actual pulses delivered using these custom beams.*

*(B) shows delivery of 4 pulses with 3.5 Gy/pulse at the regular 90 Hz on top and 4 pulses with 3.5 Gy/pulse using 4 custom beam windows at an effective 4.2 Hz.*

*(C) example of actual delivery showing three ways to deliver a total dose of 14 Gy dose within same total delivery time, but with varying cases of instantaneous dose rates: the top one showing 50 pulses with 0.28 Gy/pulse at 180 Hz, middle one showing 6 pulses with 2.3 Gy/pulse at 18 Hz, and the bottom one showing custom 2 windows of 3 pulses with 2.3 Gy/pulse at 90 Hz within each window. Customized beams are now possible and this helps us to understand biological studies of FLASH mechanism.*

## DISCUSSION

The latency parameter values for both beam-on and beam-off are similar, which shows that the actuation of the gating mechanism for both are similar in our case. They are also in the same range for all the pulse repetition frequencies we studied. The impact that a given magnitude of latency has on the different pulse repetition frequencies changes because at higher frequencies the time between pulses becomes short relative to the latency. Of note, a limitation of our study is that the specific latency parameters we found could potentially differ between linac manufacturers or even between machines from the same manufacturer. Thus, a value of our study is the framework for characterizing pulse control latency that can be applied to other machines and used to optimize pulse control across machines.

In order to gain a deeper understanding of pulse delivery with respect to respiratory gating beam holding, we developed a stochastic Monte Carlo model of the system latency, modeling the system response time as a normally distributed random variable with mean and variance as parameters that could be fit to experimental data. Monte Carlo simulations using these fitted parameters demonstrated good concordance with experimental results, indicating the utility of the model for estimating optimal parameters for pulse control.

Both the adaptive and synchronization methods, with and without making use of the underlying pulse signal of the linac respectively, could be optimized to provide highly reliable control at a pulse repetition frequency of 90 Hz. Pulse control at 180 Hz was more challenging as the latency times were larger compared to the inter-pulse interval. While the synchronization method performed at least as well as the adaptive method at 90 Hz, it performed less well at 180 Hz because in this implementation, real-time detection of the first delivered pulse was not performed in the synchronization method, so there was no ability to compensate for a missed first pulse as there was in the adaptive method. The 5.55 ms inter-pulse interval at 180 Hz was within 2 standard deviations of the beam-on latency time distribution, reducing the reliability of delivering the first intended pulse, though this was still in an acceptable range for some experiments. Adding real-time pulse detection to the synchronization method is expected to provide the best performance at the higher frequency.

We implemented custom pulse repetition frequencies or patterns beyond the standard ones preset in a linear accelerator, provided the time between adjacent pulses is an integral multiple of the fundamental period (5.55 ms). These customized beams are an interesting research topic and their biological impact with respect to FLASH is being currently studied. Our ability to understand the gating performance and overcome latency effects was essential to this capability. While reliable delivery of a single pulse proved challenging with the adaptive method, the synchronization method performed well for single pulse delivery at an underlying repetition rate of 90 Hz (Table S2 & Figure S4, Supplementary Material). Because the time from beam-on request to first delivered pulse is consistent in the synchronization method, it is also more suitable for customized pulse sequences.

The dosimetric implications of these pulse-control errors are directly determined by the dose per pulse. With dose-per-pulse values ranging from 0.5 to 4 Gy in our configuration, an individual pulse-count error corresponds to a dose uncertainty of 0.5–4 Gy – a clinically and radiobiologically significant magnitude given that the FLASH sparing effect has been reported to depend sensitively on dose above threshold values of ~5–10 Gy. For a 5-pulse delivery at 3.5 Gy/pulse (17.5 Gy total), the >99% success rate achieved with the synchronization method at 90 Hz corresponds to a mean dose uncertainty of <0.2% of prescription, whereas the 80% success rate at 180 Hz translates to an expected dose deviation of up to ~4% per delivery. This framework for translating pulse-count accuracy into dosimetric uncertainty should be applied when selecting an operating point for a given radiobiological experiment.

Our measured beam-on latency of approximately 2 ms is consistent with prior characterizations of clinical linac gating systems, which have reported latencies in the range of 50–200 ms for full-beam gating using respiratory-gating interfaces on Varian and Elekta platforms [25–29]. The substantially smaller values we observe reflect the fact that our measurement isolates the pulse-level response of the relay-driven gating path rather than the full motion-management loop, which includes surrogate acquisition, signal processing, and safety interlocks. Our values are most directly comparable to those of Shepard et al. [29], who characterized the low-level triggering latency relevant for motion management on a clinical linac. This comparison supports the interpretation that pulse-level RGS latency is dominated by the electromechanical response of the gating relay and console rather than by software-level motion-management delays.

Several limitations of this study should be acknowledged. First, our characterization was performed on a single Varian Trilogy linac at one electron energy (~18.8 MeV); latency parameters may differ between machines, between manufacturers, and potentially across service cycles on the same machine, so site-specific characterization is recommended prior to experimental use. Second, we characterized latency only at three discrete pulse repetition frequencies (60, 90, and 180 Hz); intermediate or higher PRFs were not examined. Third, the synchronization method as implemented here did not include real-time detection of the first delivered pulse, which likely limits its performance at 180 Hz; adding this feedback is expected to further improve reliability at high PRFs. Fourth, long-term stability and drift of the latency parameters over weeks-to-months of operation were not assessed. Fifth, the dose-per-pulse calibration was performed using EBT3 film cross-referenced against a Farmer chamber positioned downstream, and a complete uncertainty budget incorporating ion-recombination corrections appropriate for UHDR ionization-chamber dosimetry is beyond the scope of this methods-focused paper.

**CONCLUSION**

In this study we characterized the latency in the respiratory gating mechanism of a clinical linac for fine control of pulse delivery. We demonstrated that accounting for latency (adaptive method) and/or using the ability to read the prior information on Pulses Intended To Beam on (synchronization method) can provide high accuracy in delivering the required number of pulses for an experiment. This reliability is critical for accurate dose delivery in preclinical

FLASH research of single fraction and especially fractionated dosing regimens. The ability to generate custom pulse sequences enables more detailed exploration of the temporal dependence of biological FLASH effects.

Page 27

**Supplementary**

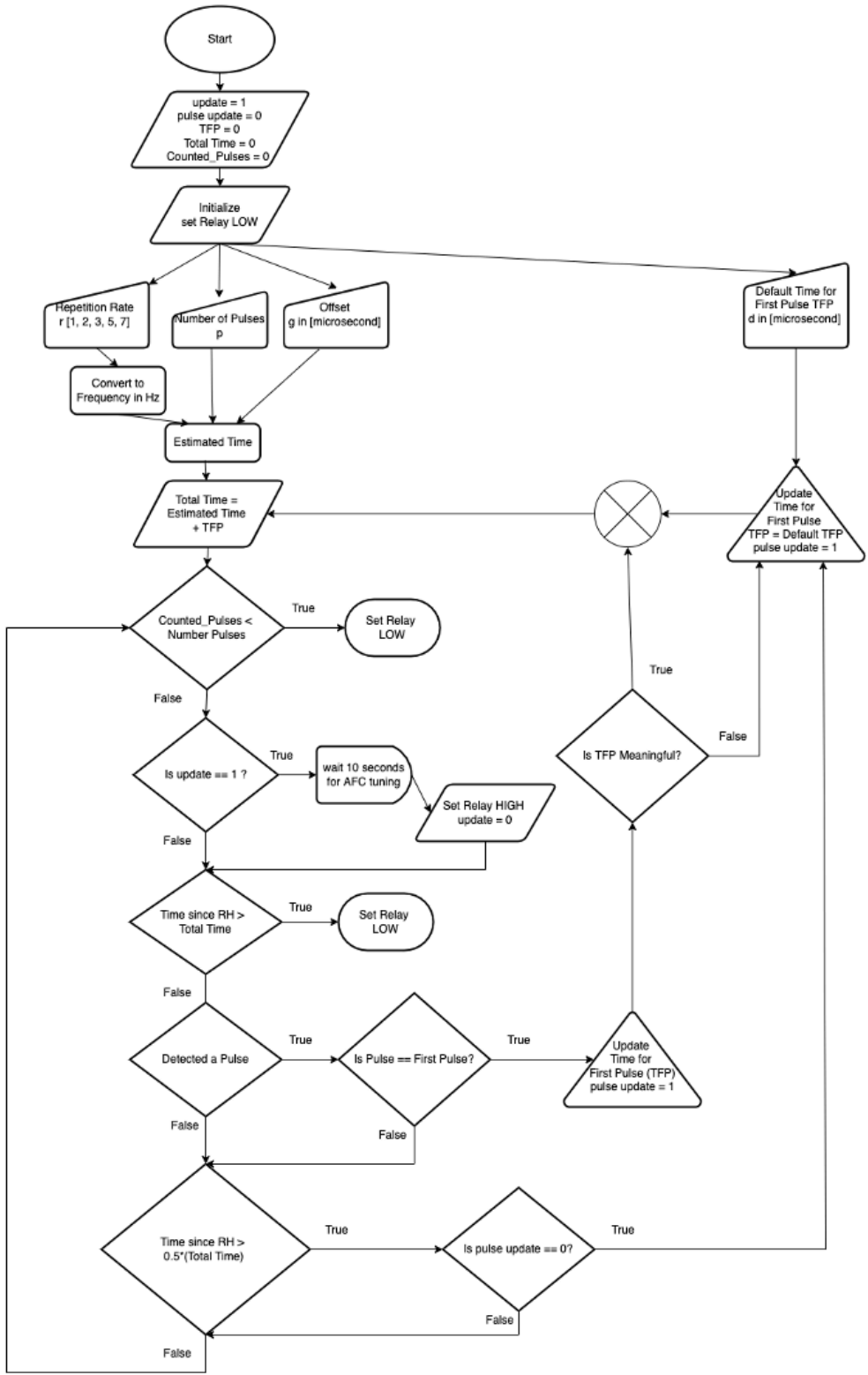

**Figure S1:** *Flowchart of the algorithm used in the adaptive method of beam control.*

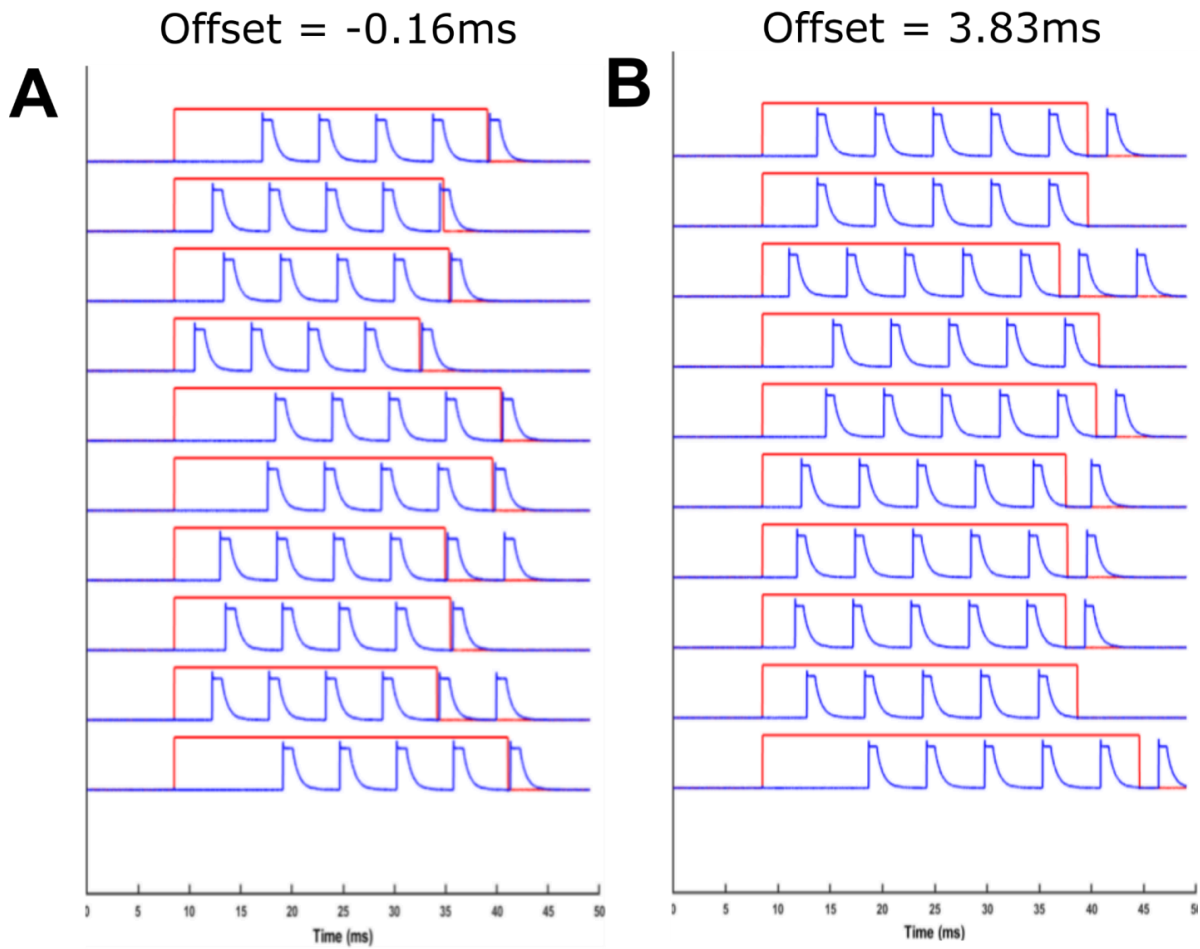

**Figure S2.** *Experimental instances of beam delivered with request of 5 pulses at 180 Hz for using the adaptive method using (A) offset of -0.16 ms, which is close to the optimal offset and (B) using a larger offset value of +3.83 ms.*

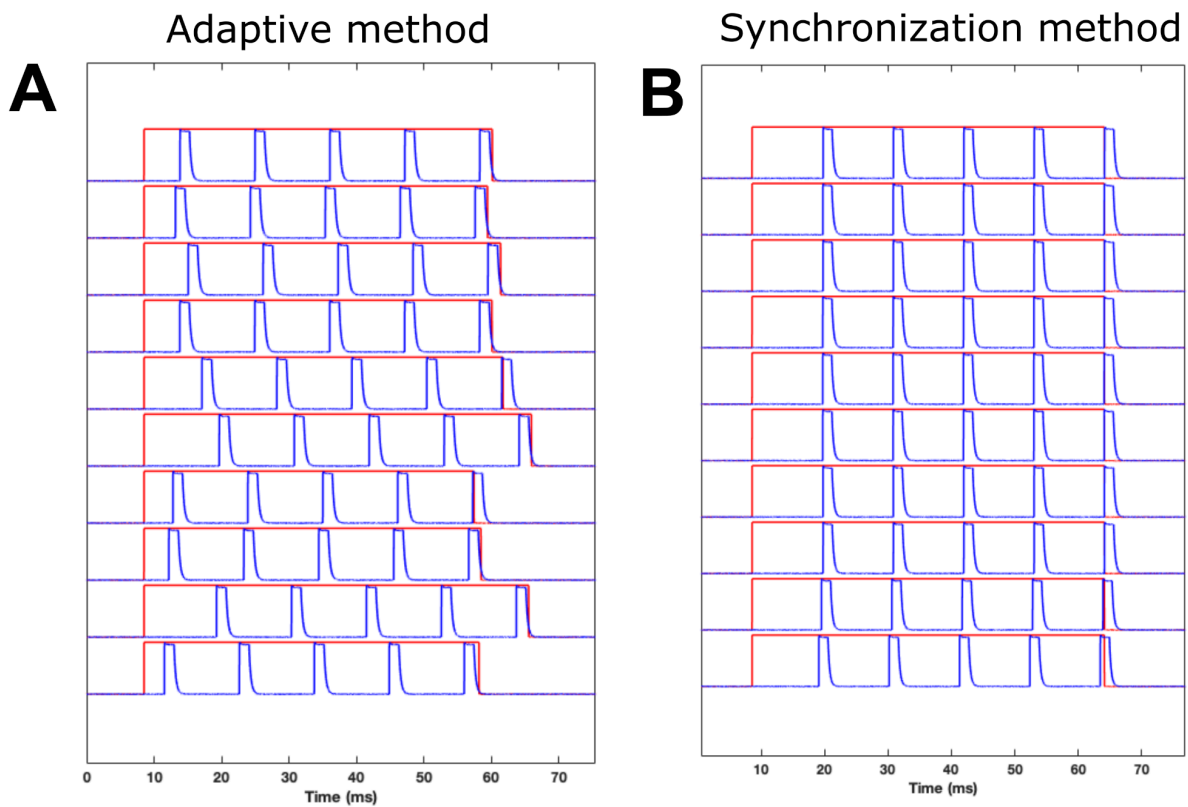

**Figure S3.** *Experimental samples for beam delivery of five pulses at 90 Hz (A) using the **adaptive method** and (B) using the **synchronization method** with the Pulses Intended To Beam on (PLSITB) signal. With the adaptive method, the total time for which the beam relay gating is set high varies depending on the actual delivery of the first pulse (which is detected in real time using the internal monitor chamber signal), while in the synchronization method the relay is set high for a fixed optimal time interval (~10 ms for the 90 Hz case) prior to the potential pulse (the time to next intended pulse [TNP]) and the total time of the relay high signal is fixed for all the cases with the same number of intended pulses. In this implementation, the synchronization method does not make use of real-time detection of the first delivered pulse.*

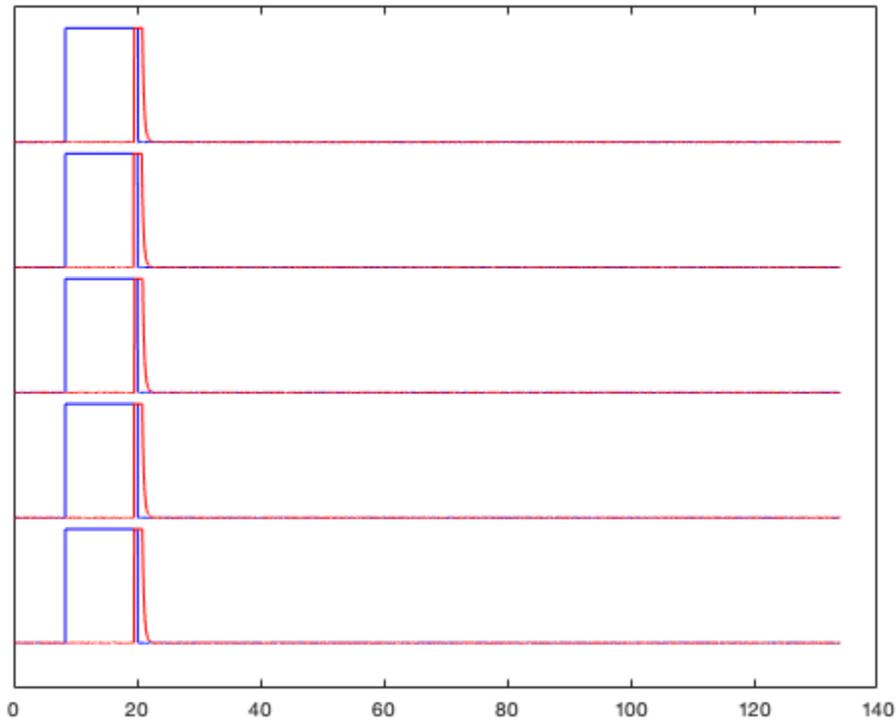

**Figure S4:** *Delivery of a single pulse is consistent at an underlying pulse repetition frequency of 90 Hz using the Synchronization method. At this frequency, the Time to Next intended Pulse (TNP) can be set long enough compared to the beam-on latency time that the reliability of delivery is nearly 100%. The relay is set low as soon as the pulse is delivered to inhibit delivery of any extra pulse.*

**Table S1:** Performance of Adaptive method for delivering 5 pulses.

| Pulse Repetition Frequency (Hz) | Offset (ms) | Total Relay High Time (ms) | Number of correct deliveries | Total number of deliveries | Percent correct |
|---|---|---|---|---|---|
| 90 | -1.16 | 51.9 | 195 | 200 | 97.5% |
| 90 | -0.16 | 52.9 | 224 | 225 | **99.5%** |
| 90 | +0.84 | 53.9 | 180 | 200 | 90% |
| 180 | -1.16 | 25.7 | 287 | 300 | 95.6% |
| 180 | -0.16 | 26.7 | 190 | 200 | **95%** |
| 180 | +0.84 | 27.7 | 174 | 200 | 87% |

.

**Table S1:** *In the Adaptive method, we tested the effect of the Offset parameter by varying its value, and tabulated the proportion of correct deliveries of the intended number of pulses (5). The optimal value of -0.16 ms corresponds well with that predicted from the measurement and modeling of latency parameters.*

**Table S2:** Performance of Synchronization method for varying number of intended pulses.

| Pulse Repetition Frequency (Hz) | Intended Number of pulses | Number of correct deliveries | Total number of deliveries | Percent correct |
|---|---|---|---|---|
| 90 | 1 | 92 | 93 | 98.9% |
| 90 | 5 | 289 | 290 | 99.6% |
| 180 | 1 | 133 | 150 | 88.6% |
| 180 | 2 | 78 | 100 | 78% |
| 180 | 3 | 80 | 100 | 80% |
| 180 | 4 | 157 | 194 | 80.9% |
| 180 | 5 | 324 | 403 | 80.5% |
| 180 | 7 | 79 | 100 | 79% |

**Table S2.** *Performance of the synchronization method with optimal offset for **varying number of intended pulses**. At a pulse repetition frequency of 90 Hz, the synchronization method performs at least as well as the adaptive method. However, at 180 Hz, the synchronization method has decreased performance compared to the adaptive method. The 5.55 ms inter-pulse interval at 180 Hz was within 2 standard deviations of the latency time distribution, reducing the reliability of delivering the first intended pulse. In this implementation, real-time detection of the first delivered pulse was not performed in the synchronization method, so there was no ability to compensate for a missed first pulse as there was in the adaptive method.*

**Table S3:** Effect of Offset on performance of Synchronization method for delivering 5 pulses.

| Pulse Repetition Frequency (Hz) | Offset (ms) | Total Relay High Time (ms) | Number of correct deliveries | Total number of deliveries | Percent correct |
|---|---|---|---|---|---|
| 90 | -1.46 | 45.9 | 63 | 98 | 64.3% |
| 90 | -0.16 | 47.2 | 100 | 100 | **100%** |
| 90 | +0.84 | 48.2 | 74 | 75 | 98.6% |
| 180 | -1.46 | 23.7 | 49 | 100 | 49% |
| 180 | -0.16 | 25 | 164 | 203 | **80.8%** |
| 180 | +0.84 | 26 | 77 | 100 | 77% |

**Table S3.** *Table showing the effect of Offset at 90 Hz and 180 Hz pulse repetition frequencies for delivering 5 pulses using the synchronization method after optimizing the TNP parameter. The value of optimal Offset is crucial as the proportion of correct deliveries decreases at higher and lower values.*